\documentclass{revtex4}

\usepackage{graphicx}

\date{\today}

\begin{document}

\title{Vibrational energy relaxation (VER) of isotopically labeled amide I modes 
in cytochrome c: Theoretical investigation of VER rates and pathways}

\author{Hiroshi Fujisaki}\email{fujisaki@theochem.uni-frankfurt.de} 
\affiliation{
Department of Chemistry,
Boston University, 590 Commonwealth Avenue,
Boston,  Massachusetts 02215, USA
}
\affiliation{
Institute of Physical and Theoretical Chemistry,
J.W. Goethe University, Max-von-Laue-Str. 7,
60438 Frankfurt am Main, Germany
}
\author{John E. Straub}\email{straub@bu.edu} 
\affiliation{
Department of Chemistry,
Boston University, 590 Commonwealth Avenue,
Boston,  Massachusetts 02215, USA
}
\affiliation{
Department of Chemistry and Biochemistry,
Montana State University, Bozeman, Montana 59717, USA
}

\begin{abstract}
Using a time-dependent perturbation theory, 
vibrational energy relaxation (VER) 
of isotopically labeled amide I modes in cytochrome c 
solvated with water is investigated.  
Contributions to the VER are decomposed into 
two contributions from the protein and water.
The VER pathways are visualized using radial and 
angular excitation functions for resonant normal modes.
Key differences of VER among different amide I modes 
are demonstrated, leading to a 
detailed picture of the spatial anisotropy of the VER.
The results support the experimental 
observation that amide I 
modes in proteins relax with sub picosecond timescales, 
while the relaxation mechanism 
turns out to be sensitive to the environment of the amide I mode.

\end{abstract}





\maketitle

\section{Introduction}

Amide I vibrational modes in proteins or peptides 
are localized around the peptide backbone and can be 
a sensitive probe of protein structure and dynamics. Mainly localized around 
the CO bond of the backbone with large oscillator strength,
the amide I modes detected using infrared (IR) spectroscopy
have been studied for a variety of proteins and peptides \cite{KB86,TT92,BZ02,SAD03,GHG04}.  
Recently 2D-IR spectroscopy \cite{TM93,ZH01,WH02} has been utilized to 
decipher (anharmonic) coupling between vibrational modes including 
the amide I mode \cite{WH06,Tokmakoff06}.
The interpretation of the associated spectra is not necessarily simple, 
and will benefit from the development of increasingly accurate 
theoretical models \cite{IT06,NTM07,Torii07}.

Combining MD simulations with {\it ab initio} calculations, 
several theoretical groups have recently devised sophisticated methods to 
characterize 
the effect of inhomogeneity on the dephasing time, $T_2$, 
of the amide I mode, using small peptides 
in water \cite{HHLKKC04,Skinnergroup,ZAHM06,HJZM05,Stock06}. 
Anharmonic frequency calculations have been performed for 
a peptide-like molecule, N-methylacetamide (NMA), 
in vacuum \cite{GCG02,BS06,KB07,FYHS07} 
and in water \cite{HJZM05}.
For the amide I mode of NMA in water, 
theoretical investigations of VER have been carried out using a quasi-classical 
method \cite{NS03} and a quantum mechanical perturbation method \cite{FZS06}. 
However, while there have been many experimental studies \cite{HLH98,ZAH01,XMHA00} 
of  
vibrational energy relaxation (VER) of amide I modes in proteins, 
related to $T_1$, 
there are relatively few theoretical studies.

In this work, we study
the VER properties of amide I modes in a protein, cytochrome c, in water,
and clarify the VER pathways using 
quantum-mechanical time-dependent perturbation theory \cite{FZS06}. 
This is a continuation of our previous work on VER of a CD stretching 
mode in cytochrome c \cite{FBS05,Cremeens06}.
The amide I modes studied were isotopically labeled 
according to IR experiment \cite{MKFKAZ04,Arkin06}. 
We decompose the VER rate into two components: 
one from the protein, and the other from water (solvent).    
We identify the resonant modes that contribute most significantly 
to the VER rate, 
and introduce distribution-like functions for those modes to visualize 
the spatial anisotropy of VER as pathways in 3-dimensional space. 

This paper is organized as follows.
In Sec.~\ref{sec:methods}, 
our time-dependent perturbation method is briefly described,
and several methods to visualize the VER pathways in protein 
and water are proposed.
In Sec.~\ref{sec:results}, 
we examine the results of VER rates by comparing with 
the Maradudin-Fein formula, SASA, localization length, 
and experimental results. Furthermore, we mention 
the anisotropy of VER from excited amide I modes 
into protein as well as into water.
In Sec.~\ref{sec:summary}, we summarize and discuss 
the future extension of our work.

\section{Methods}
\label{sec:methods}

\subsection{VER rate formula}

We briefly summarize the quantum mechanical perturbation theory
employed in this work. 
We assume that normal mode coordinates $q_{\alpha}$ 
provide a good description of the system dynamics for the observables 
studied. We further assume that the potential energy function 
can be Taylor expanded up to the 3rd and 4th order 
anharmonic terms with respect to normal mode coordinates 
including only the relaxing mode $q_S$ \cite{FZS06}. 
\begin{eqnarray}
H &=& {\cal H}_S +{\cal H}_B
-q_S \delta {\cal F} +q_S^2 \delta {\cal G}
\label{eq:ham}
\\
{\cal H}_S &=& \frac{p_S^2}{2}+V(q_S)
\\
{\cal H}_B &=& \sum_{\alpha} \frac{p_{\alpha}^2}{2}+ \frac{\omega_{\alpha}^2 q_{\alpha}^2}{2}
\\
\delta {\cal F}
&=&
\sum_{\alpha,\beta} C_{S \alpha \beta} 
(q_{\alpha} q_{\beta} - \langle q_{\alpha} q_{\beta} \rangle)
\\
\delta {\cal G}
&=&
\sum_{\alpha,\beta} C_{S S\alpha \beta} 
(q_{\alpha} q_{\beta} - \langle q_{\alpha} q_{\beta} \rangle)
+\sum_{\alpha} C_{SS \alpha} q_{\alpha}
\end{eqnarray}
where ${\cal H}_S$ (${\cal H}_B$) is the system (bath) Hamiltonian,
and $C_{S \alpha \beta}$ ($C_{SS \alpha \beta}$) are the 3rd (4th) order  
coupling terms.
(This is related to the three-mode representation of 
a quartic force field \cite{FYHS07,YHTSG04}.)
From the von Neumann-Liouville equation, 
a reduced density matrix for the relaxing mode
is derived using the time-dependent perturbation theory 
after tracing over the bath degrees of freedom. 
(A similar result has been derived from the path integral 
formulation of quantum mechanics by Okazaki and coworkers \cite{SO98}.)
The Markov approximation is usually employed to 
derive a simplified Bloch-Redfield type 
equation after introducing the density of states for 
the bath. However, to describe the initial stage of 
the quantum dynamics, we do not invoke the Markov 
approximation. Using this approach, we are able to avoid 
assumptions related to the lifetimes of the vibrational modes of the 
bath --- 
the so-called ``linewidth problem'' \cite{FBS05} used in 
the Markov approximation. 

When the relaxing mode is excited to the $v=1$ state,
the VER is described by the decay of the reduced density 
matrix element $(\rho_S)_{11}(t)$. 
We define the temporal VER rate as $R(t)=d (\rho_S)_{11}(t)/dt$,
which is approximately written as \cite{FZS06}
\begin{eqnarray}
R(t) 
&\simeq&
\frac{2}{\hbar^2}
\sum_{\alpha,\beta} 
\left[
C_{--}^{\alpha \beta}  
\dot{u}_t(\tilde{\omega}_S-\omega_{\alpha}-\omega_{\beta})
+C_{++}^{\alpha \beta}  
\dot{u}_t(\tilde{\omega}_S+\omega_{\alpha}+\omega_{\beta})
\right.
\nonumber
\\
&&
\left.
+C_{+-}^{\alpha \beta}  
\dot{u}_t(\tilde{\omega}_S-\omega_{\alpha}+\omega_{\beta})
\right]
\equiv 
\sum_{\alpha,\beta}R^{\alpha \beta}(t)
\label{eq:VERrate}
\end{eqnarray}
where $\dot{u}_t(\Omega) = \sin \Omega t/\Omega$, $\tilde{\omega}_S$ is the 
anharmonicity-corrected system frequency, 
$\omega_{\alpha}$ is the bath mode (harmonic) frequency, and the 
coefficients $C_{++}^{\alpha \beta},C_{+-}^{\alpha \beta},C_{--}^{\alpha \beta}$ 
are derived from the nonlinear coupling constants 
$C_{S \alpha \beta}$ and $C_{SS \alpha \beta}$ \cite{FZS06}.
$R^{\alpha \beta}(t)$ is each component of the VER rate corresponding 
to modes $\alpha$ and $\beta$.
We note  that the first term in Eq.~(\ref{eq:VERrate}) dominates in the 
formula because of the resonance condition, i.e. 
$\dot{u}_t(\Omega)$ becomes large when 
$\tilde{\omega}_S-\omega_{\alpha}-\omega_{\beta} \simeq 0$.  

Because of non-Markovian properties, 
$R(t)$ is not necessarily a constant. 
If the Markov assumption holds, $R(t)$ 
becomes a {\it rate constant}.
In the case of the amide I modes in cytochrome c, 
we have confirmed that a ``VER rate'' 
can be defined as a time average of 
$R(t)$ {\it after a certain transient time} $t_{tr}$
\begin{equation}
\bar{R}
\equiv 
\frac{1}{T-t_{tr}}
\int_{t_{tr}}^{T} R(t) dt 
=\sum_{\alpha \beta} \bar{R}^{\alpha \beta}.
\label{eq:VERrate2}
\end{equation}
using $T \simeq 1.0$ ps and  $t_{tr} \simeq 0.5$ ps. 
With this procedure,
the averaged VER time is obtained as $T_1 \simeq 1/\bar{R}$. 
The final rate formula is similar to the 
Maradudin-Fein (MF) formula \cite{FBS05,Leitner05}, and both formulas 
lead to a rate. An essential difference is that we avoid invoking 
the Markov approximation and introducing the phenomenological 
linewidth parameter.
However, the MF formula has been widely employed in the literature
\cite{FBS05,Leitner05}, and it would be interesting to examine the 
usefulness of the MF formula based on our formula.
Hence we introduce the MF rate as 
\begin{equation}
R^{MF}=\frac{2}{\hbar^2} \sum_{\alpha,\beta}
C_{--}^{\alpha \beta} \frac{\delta}{\delta^2 + (\tilde{\omega}_S-\omega_{\alpha}-\omega_{\beta})^2}
\label{eq:MF}
\end{equation}
where $\delta$ is the linewidth parameter which will be discussed below.

\subsection{Analysis of VER pathways}

As shown in Eq.~(\ref{eq:VERrate2}),
the VER rate $\bar{R}$ can be decomposed into components $\bar{R}^{\alpha \beta}$.
This result allows us to examine 
which modes (combination) contribute significantly to the VER rate. 
To develop a detailed picture of VER in a complex molecular system,
we utilize the normal mode eigenvectors $U^{\alpha}_k$ which satisfy
\begin{equation}
\sum_{kl} U^{\alpha}_k F_{kl} U^{\beta}_l = \omega_{\alpha}^2 \delta_{\alpha \beta}
\end{equation} 
where $\alpha$ ($k$) represents the normal mode index (Cartesian coordinate 
index) and $F_{kl}$ is the mass-weighted Hessian matrix.

We first introduce the localization length of the normal mode \cite{SS01,DK05}
\begin{equation}
L^{\alpha} = \left[ \sum_{k} |U_k^{\alpha}|^4 \right]^{-1}
\label{eq:IPR}
\end{equation}
which measures how a normal mode extends in a system in terms of 
the number of atoms. This measure (also called inverse participation ratio) 
is used to quantify the localization 
of the amide I modes.

We are interested in evaluating contributions from protein and water to 
the value and mechanism of VER.
We assign modes to water (protein) if the condition $\sum_{k \in {\rm water}}
|U^{\alpha}_k|^2 > 0.5$ 
($\sum_{k \in {\rm protein}}
|U^{\alpha}_k|^2 > 0.5$)
is satisfied. Here $k \in {\rm water (protein)}$ means that the summation 
is taken only for water (protein) degrees of freedom.
Hence $\bar{R}^{\alpha \beta}$ can be decomposed into the
(1) protein-protein contribution (if $\alpha$ and $\beta \in {\rm protein}$),
(2) protein-water contribution (if $\alpha (\beta) \in {\rm protein}$ and 
$\beta (\alpha) \in {\rm protein}$),
and
(3) water-water contribution (if $\alpha$ and $\beta \in {\rm water}$).

\begin{figure}[h]
\hfill
\begin{center}
\includegraphics[scale=0.8]{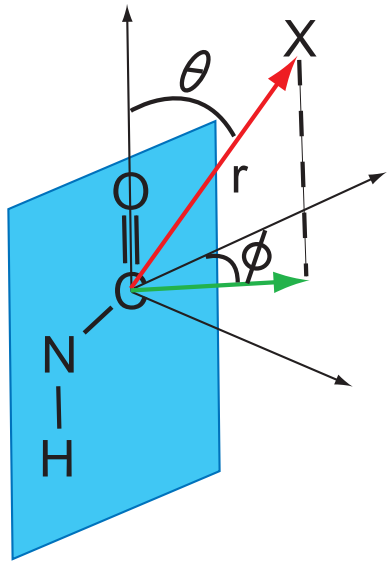}
\end{center}
\caption{
The polar coordinates $(r,\theta,\phi)$ for an atom $X$
used to calculate the radial and angular 
excitation functions of a normal mode. 
The origin of the coordinate system 
is the $C_{\beta}$ atom of the 
amide I mode.
The sheet represents the amide plane.
}
\label{fig:geometry}
\end{figure}

We next introduce radial and angular {\it excitation distribution functions} 
for each normal mode $\alpha$ as 
\begin{eqnarray}
g^{\alpha}(r) 
&=&
\sum_{k \in \Delta \Gamma} |U^{\alpha}_k|^2
\\
h^{\alpha}(\theta,\phi) 
&=&
\sum_{k \in \Delta \Omega} |U^{\alpha}_k|^2
\end{eqnarray}
where $\Delta \Gamma$ and $\Delta \Omega$ are the bins for the radial direction 
(whose median is $r$)
and the bins for the cubic angle (whose median is $\theta,\phi$), respectively
(see Fig.~\ref{fig:geometry}). 
The summation is only taken over these bins.
We refer to  
$g^{\alpha}(r)$ and $h^{\alpha}(\theta,\phi)$
as excitation distribution functions because 
they represent coarse-grained spatial information 
about the excitation of a normal mode. 

Note that these are different from the ordinary distribution functions 
(the absolute value does not have physical meaning),
and are not directly related to experimental observables (such as neutron
scattering functions). 
These functions are specifically introduced for visualizing the 
VER pathways. Other approaches are possible. 
For example, Dijkstra and Knoester visualized 
normal modes of $\beta$-hairpin and $\beta$-sheet peptides using 
a color map in their analysis of IR and 2D-IR spectra \cite{DK05}.

We define the average excitation distribution functions 
for resonant modes as 
\begin{eqnarray}
\bar{g}(r) 
&=&
\frac{1}{N_{\rm res}}
\sum_{\alpha \in {\rm res}} 
g^{\alpha}(r) 
\label{eq:ex1}
\\
\bar{h}(\theta,\phi) 
&=&
\frac{1}{N_{\rm res}}
\sum_{\alpha \in {\rm res}} 
h^{\alpha}(\theta,\phi) 
\label{eq:ex2}
\end{eqnarray}
where $\alpha \in {\rm res}$ indicates that the summation is 
taken only for resonant modes, meaning
those modes with $\bar{R}^{\alpha \beta} > R_{th}$. 
$N_{\rm res}$ is the number of such resonant modes.
We took $R_{th} \simeq 0.04$ ps$^{-1}$ 
for purposes of illustration.
As indicated in Eq.~(\ref{eq:VERrate}), 
when a system mode $S$ is excited, 
the excess energy mainly flows to 
resonant (bath) modes $\alpha$ and $\beta$ via Fermi resonance 
($\tilde{\omega}_S - \omega_{\alpha} -\omega_{\beta} \simeq 0$).
We can explore the spatial content of such 
resonant modes using the excitation distribution functions.

Sagnella and Straub introduced another way to visualize  
anisotropy of energy flow in a protein \cite{SS01}, and the differences are 
(1) we consider quantum mechanical energy flow whereas their 
calculation is classical, 
and (2) we employ the resonant normal modes whereas they 
used excess kinetic energies as the definition of the energy flow pathway.
Okamoto and Nagaoka utilized an approach derived from hydrodynamics 
to visualize the anisotropic energy flow from a diatomic 
molecule in water \cite{ON05}, though their work is also based on 
classical mechanics.
Much similar to the present work is that by Mikami and Okazaki
who devised a strategy for analyzing anisotropy of 
energy flow using quantum mechanics \cite{MO03}, but the method 
is tailored for VER problems of a diatomic molecule.
In this work, we take the naive approach described above 
to visualize resonant normal modes.

\subsection{Numerical procedure}

We used CHARMM \cite{CHARMM} to construct and simulate 
the system consisting of cytochrome c and solvent water. 
CHARMM facilities were utilized to calculate dynamics, normal modes, 
solvent accessible surface area (SASA) etc., and 
the computational details have been presented elsewhere \cite{BS03}.
After equilibration, a 100 ps trajectory was generated and configurations 
were saved every 1 ps.
From the 100 sample configurations, we calculated the VER rate 
using Eq.~(\ref{eq:VERrate2}) and the results were averaged.

Application of our formula starts from the decomposition of 
the system Hamiltonian using normal modes as in Eq.~(\ref{eq:ham}).
However, computation of normal modes for the 
full system is prohibitive. 
Hence we take a reduced system strategy:
We deleted all atoms except those in 
a spherical region $r < R_c$ (see Fig.~\ref{fig:geometry})
where $R_c =12$ \AA~  \cite{FZS06}. 
For 
the reduced system,
we computed the anharmonic coefficients $C_{S \alpha \beta}, C_{S S \alpha \beta}$  
using a finite difference approximation \cite{FBS05}
and the normal mode frequencies $\omega_{\alpha}$ 
using instantaneous normal mode analysis \cite{NS03,Stratt95} (imaginary 
frequencies are neglected). 
The summation in Eq.~(\ref{eq:VERrate2}) was taken only 
for the bath modes with $\omega_{\alpha} > \omega_c$,
where we used $\omega_c=100$ cm$^{-1}$ as the frequency cutoff. 
This is a reasonable approximation, as we have observed that
 the low frequency modes do not contribute significantly to the VER rate.

In the CHARMM program, 
we isotopically labeled {\it one} CO bond in a residue 
as $^{13}$C$=^{18}$O \cite{Arkin06,MKFKAZ04} and 
we examined four different cases as described below. 
It was expected that isotopical labeling should localize 
the amide I mode, but this is not necessarily the case as shown below.

\section{Results and discussions}
\label{sec:results}

\subsection{Protein and water contribution to VER rate}

\begin{figure}[h]
\hfill
\begin{center}
\begin{minipage}{.42\linewidth}
\includegraphics[scale=0.4]{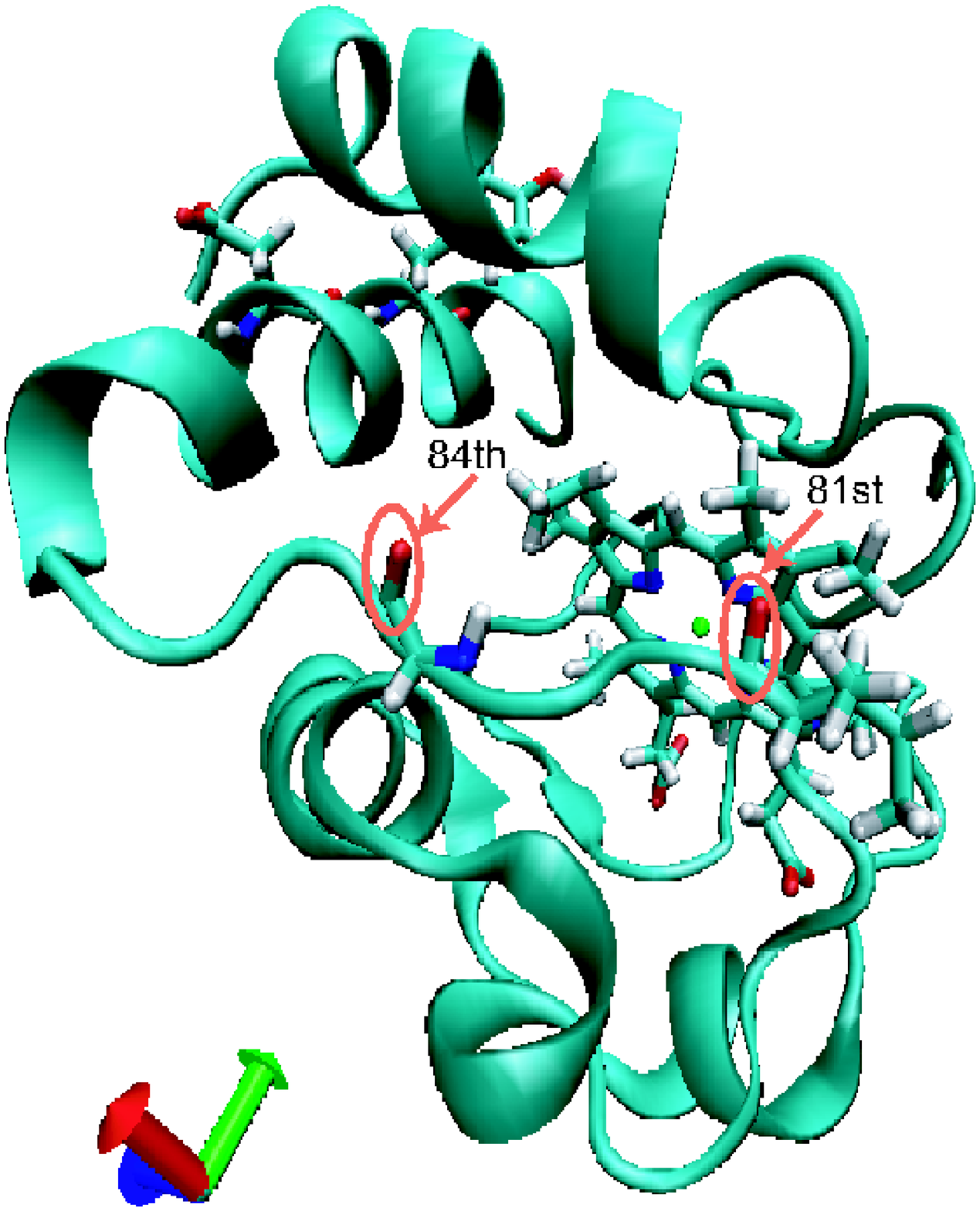}
\end{minipage}
\hspace{1cm}
\begin{minipage}{.42\linewidth}
\includegraphics[scale=0.4]{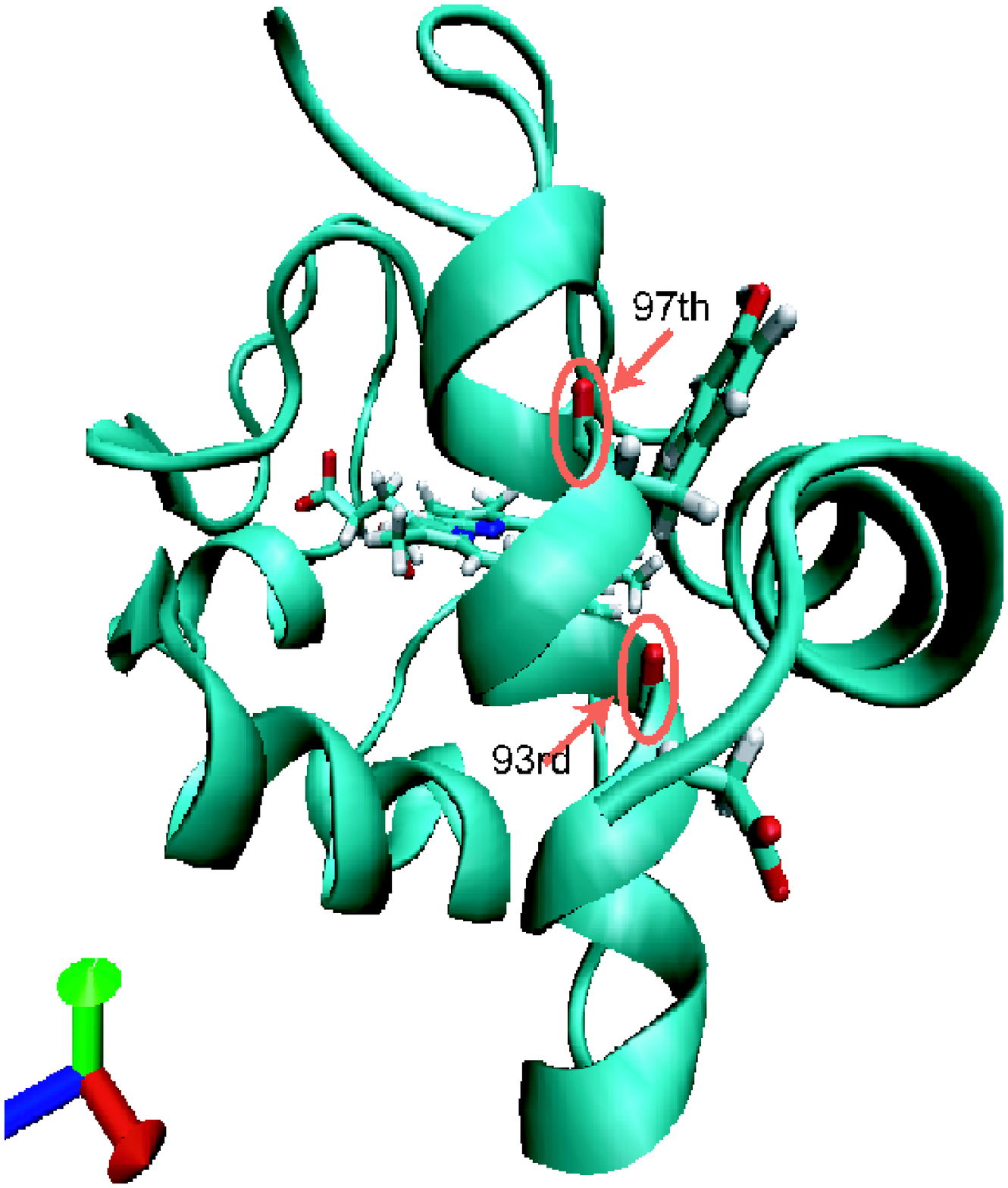}
\end{minipage}
\end{center}
\caption{
Left: 
81st and 84th residues in cytochrome c in a loop region. 
Right: 
93rd and 97th residue in cytochrome c in a $\alpha$ helical region. 
The cartoon represents the protein, 
and the licorice the four residues.
(The water molecules are excluded for simplicity.)
VMD (Visual Molecular Dynamics)
was used to generate these figures \cite{VMD}.
}
\label{fig:residue}
\end{figure}

 We isotopically labeled four residues: 
81st, 84th, 93rd, and 97th (see Fig.~\ref{fig:residue}).
The first two belong to a loop region of cytochrome c 
whereas the latter two to a $\alpha$ helical region.
By examing SASA (Table \ref{table:VER}), 
we see that the CO atoms of the 84th residue is most exposed to solvent 
whereas those of the 97th residue is most buried in the protein.

\begin{table}[htbp]
\hfill
\caption{
Protein and water contributions to the VER rate for the amide I modes of 
four residues in cytochrome c in water.
The total rate is $\bar{R}$, and the protein-protein, 
protein-water, water-water contribution are denoted as 
 $\bar{R}_{pp}, \bar{R}_{pw}$ and $\bar{R}_{ww}$, respectively.
The value derived from the Maradudin-Fein formula is 
denoted as $R^{MF}$. The units are in ps$^{-1}$ for the VER rate. 
The inverse participation ratio (localization length) of 
the amide I modes and solvent accessible surface area (SASA)
for the CO atoms are also shown. The units for SASA is \AA$^2$.
The number in the parenthesis
represents the standard error of the mean.
}
\begin{center}
\begin{tabular}{c|c|c|c|c|c|c|c}
Residue & $\bar{R}$ &   $\bar{R}_{pp}$ &  $\bar{R}_{pw}$ & $\bar{R}_{ww}$ & $R^{MF}$ & IPR & SASA(CO) \\ \hline \hline           
81st   &  2.29 (0.03) &  0.95 (0.02) &  1.21 (0.02)  & 0.13 (0.00) & 2.24 (0.03) & 9.54 (1.03) & $\sim$ 80 \\ \hline 

84th   &  2.36 (0.03) &  0.99 (0.02) &  1.31 (0.03)  & 0.06 (0.00) & 2.29 (0.04) & 8.26 (0.95) & $\sim$ 100 \\ \hline 
93rd   &  2.27 (0.03) &  1.18 (0.02) &  1.05 (0.02)  & 0.04 (0.00) & 2.24 (0.03) & 3.92 (0.03) & $\sim$ 90 \\ \hline           
97th   &  1.75 (0.03) &  1.46 (0.03) &  0.30 (0.01)  & 0.00 (0.00) & 1.82 (0.03) & 4.71 (0.13) & $\sim$ 70 \\ \hline            
\end{tabular}
\end{center}
\label{table:VER}
\end{table}

In Table \ref{table:VER},
we observe that the VER rates are $\simeq 2$ ps$^{-1}$ ($T_1 \simeq 0.5$ ps) 
for all the cases. 
This is similar to the VER rate of the amide I mode of
 N-methylacetamide in heavy water 
($\simeq 2.0$ ps$^{-1}$) \cite{FZS06}
though  the 97th residue has a slightly slower rate.
It is interesting to note that the MF formula Eq.~(\ref{eq:MF}) 
gives rather ``accurate'' values with an appropriate linewidth parameter 
$\delta= 3$ cm$^{-1}$.
This result seems to validate the use of the MF formula 
and to indicate that the MF formula is sufficient to 
describe VER. 
However, note that an appropriate value of the linewidth parameter 
$\delta$ can not be determined in advance.

Looking into the contributions from protein and water,
we observe that the protein-water contribution is significantly 
less for the 97th residue compared to other residues 
(81st, 84th, and 93rd).
This is because the water less surrounds the CO bond in this case 
(see Fig.~\ref{fig:residue}) as quantified by 
SASA for the CO bond (see Table \ref{table:VER}). 
There seems to be a correlation between SASA and $\bar{R}^{pw}$.
However, as shown below, the water motion 8 \AA \, away from the 
CO bond can be involved in the VER processes. Hence the 
interpretation is not so simple.

We also investigate this point in terms of 
the localization length Eq.~(\ref{eq:IPR}) for the amide I modes.
The amide I mode is expected to be localized on a CO bond 
only but this is necessarily the case as shown in Table \ref{table:VER}.
For the four residues we examined here, the amide I mode extends 
over from 4 to 10 atoms. For more delocalized modes,
the protein-water or water-water contributions are expected to be 
large because there are more contact with water.
However, this expectation is not validated in a strict sense 
as shown in Table \ref{table:VER}.
We know that the Fermi resonance parameter \cite{FYHS07,Cremeens06}
can strictly describe the VER pathways in molecules, but 
for the present system it is difficult to interpret the result based on 
some chemical intuition.

Note that the protein-protein contributions
for the 81st, 84th, and 93rd residues are smaller than that for the 97th residue. 
However, due to the protein-water contribution, the VER 
rates of the former three amide I modes are faster than the latter.
We conclude that 
VER of the former amide I modes 
is protein-mediated, whereas that of
the latter residue is water-mediated,
which should have some experimental consequences as mentioned below.


How do our calculated VER rates obtained in this theoretical study
compare with experimentally derived values? Hochstrasser and coworkers obtained 
sub-picosecond VER time scales for amide I modes in several proteins \cite{HLH98,ZAH01}.
Zanni and coworkers also observed sub-picosecond VER time scales for 
isotopically labeled amide I modes in 
a membrane protein \cite{MKFKAZ04}.
The experimentally derived VER times are comparable to our 
theoretical results (sub picosecond).
However note that although the total VER rates are similar, 
the physical mechanism is sensitive to the specific structural 
environment of the amide I mode as discussed above (Table \ref{table:VER}).
Using multidimensional spectroscopy \cite{ZH01,WH02,HLH98,ZAH01,MKFKAZ04}, it should be 
possible to discriminate the protein and water 
contributions to VER separately in experiment.

We have recently studied the solvent effects on heme 
cooling in myoglobin \cite{ZFS07}. We found that 
the mechanism for VER is dependent on resonance as well as structural 
anisotropy that modulates mode coupling.
It will be interesting 
to investigate the solvent effects on VER of amide I modes 
by changing the nature of the solvent
within the quantum mechanical perturbation theory
(while the previous work \cite{ZFS07} is based on classical mechanics).


\subsection{Visualization of VER pathways}

\begin{figure}[h]
\begin{minipage}{.42\linewidth}
\includegraphics[scale=1.0]{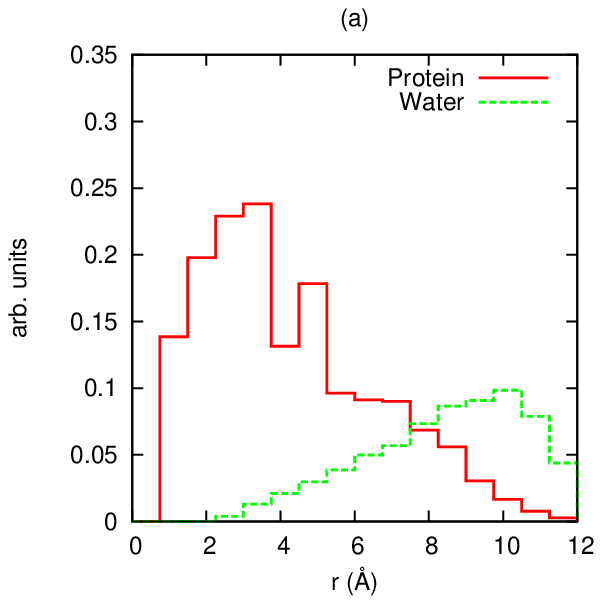}
\end{minipage}
\hspace{1cm}
\begin{minipage}{.42\linewidth}
\includegraphics[scale=1.0]{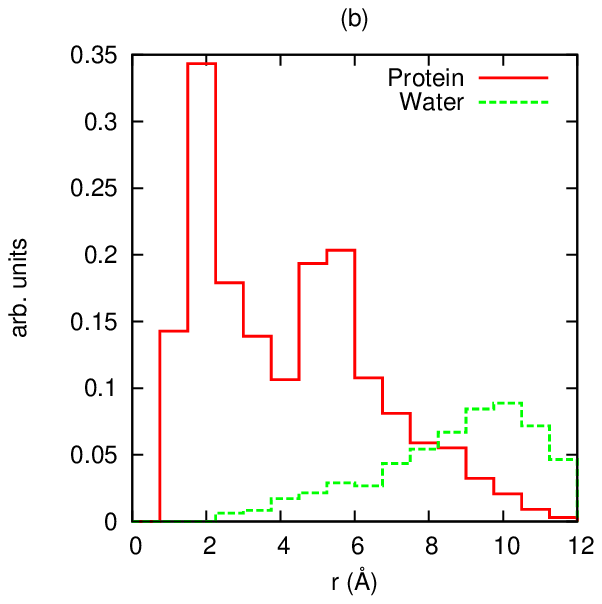}
\end{minipage}
\begin{minipage}{.42\linewidth}
\includegraphics[scale=1.0]{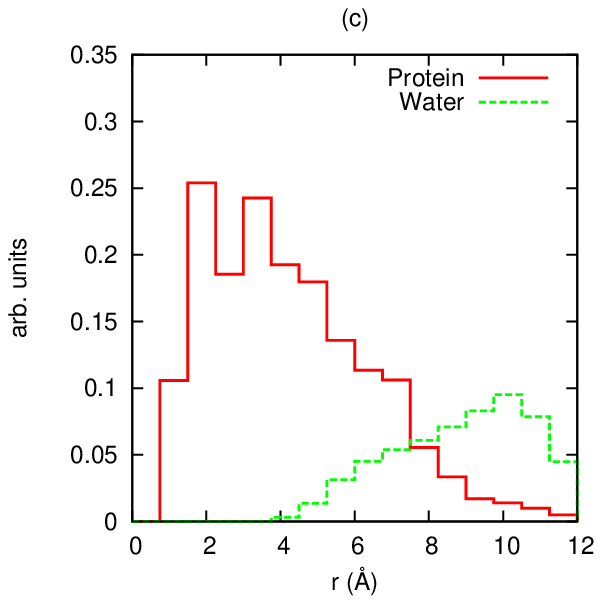}
\end{minipage}
\hspace{1cm}
\begin{minipage}{.42\linewidth}
\includegraphics[scale=1.0]{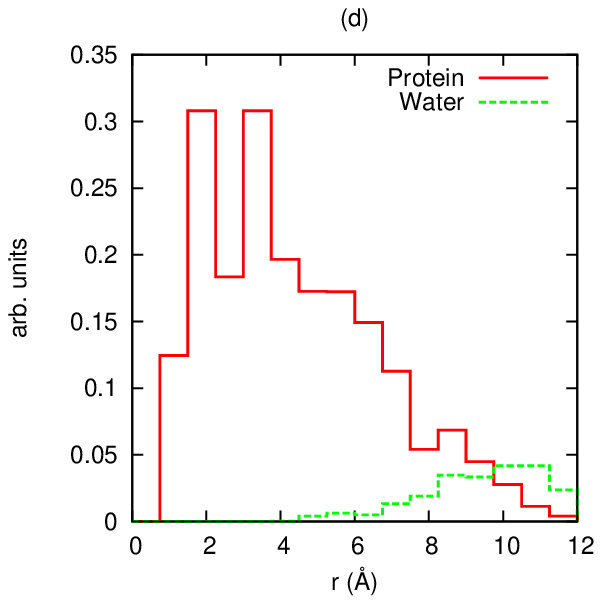}
\end{minipage}
\vspace{1cm}
\caption{
Radial excitation functions for the resonant normal modes
of protein and water 
for the (a)  81st, (b) 84th, (c) 93rd, and (d) 97th residues, 
represented in arbitrary units.
}
\label{fig:dist1}
\end{figure}

We have used the previously defined excitation functions Eqs.~(\ref{eq:ex1}) and 
(\ref{eq:ex2}) to scrutinize the spatial anisotropy of the VER pathways.
The radial excitation function in Fig.~\ref{fig:dist1} 
shows each (protein or water) contribution to VER pathways along
the radial direction.
The peaks around $r \simeq 2$ \AA \, correspond 
to the nearest neighbor intramolecular pathways 
involving the motion of the NH bond (see Fig.~\ref{fig:geometry}).
Since the density of the protein has a peak around
$r \simeq 6$ \AA \,,
the radial excitation function also has 
a peak or shoulder around at the same location.

Looking at the water contributions,
we found that the excitation functions
have significant values at large distances ($>8$ \AA),
indicating that VER is mediated by distant water molecules.  
On the other hand, Mikami and Okazaki 
found that only the water molecules 
of the first solvation shell contribute to VER in the 
case of a solvated diatomic molecule \cite{MO03}.
We note that there are significant differences 
between the relaxation of an amide I mode in a protein 
and of a solvated diatomic molecule.
For VER (or energy transfer) to occur, 
the overlap of resonant normal modes is needed \cite{MMK00}. 
In the case of a diatomic molecule, 
the stretching mode is, by definition, localized around 
the bond, and mainly couples to the neighboring water molecules.
However, in the case of amide I modes in solvated proteins,
the amide I modes are delocalized as shown in Table \ref{table:VER}. 
Many resonant modes couple to such amide I modes, which 
can contain delocalized (collective) water motions.
Further theoretical and experimental studies are needed 
to clarify the role of such delocalized modes 
on VER in proteins.

\begin{figure}[h]
\begin{center}
\begin{minipage}{.42\linewidth}
\includegraphics[scale=1.2]{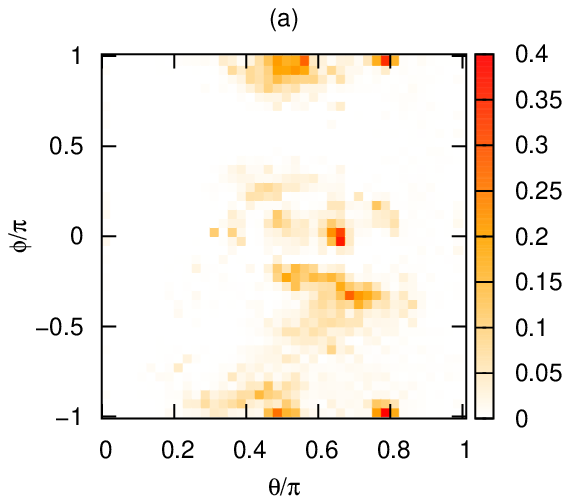}
\end{minipage}
\hspace{1cm}
\begin{minipage}{.42\linewidth}
\includegraphics[scale=1.2]{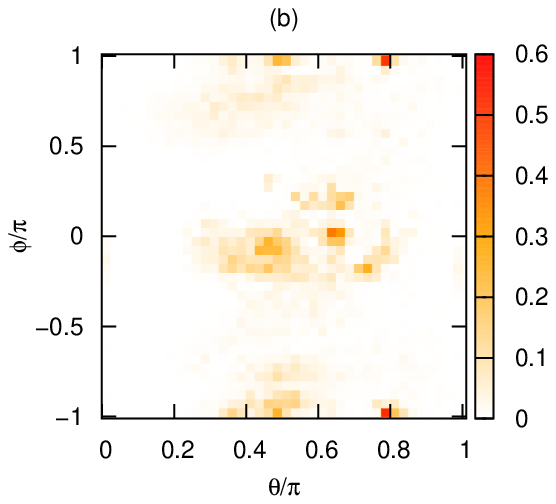}
\end{minipage}
\begin{minipage}{.42\linewidth}
\includegraphics[scale=1.2]{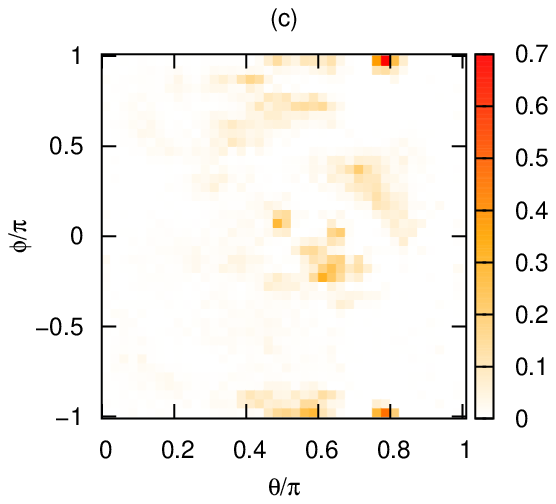}
\end{minipage}
\hspace{1cm}
\begin{minipage}{.42\linewidth}
\includegraphics[scale=1.2]{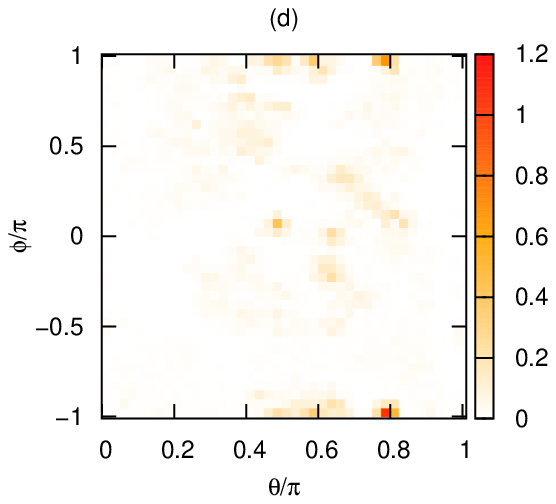}
\end{minipage}
\end{center}
\caption{
Angular excitation functions for the resonant normal modes
of the protein 
for the (a)  81st, (b) 84th, (c) 93rd, and (d) 97th residues, 
represented in arbitrary units.
}
\label{fig:dist2-pro}
\end{figure}

\begin{figure}[h]
\begin{center}
\begin{minipage}{.42\linewidth}
\includegraphics[scale=1.2]{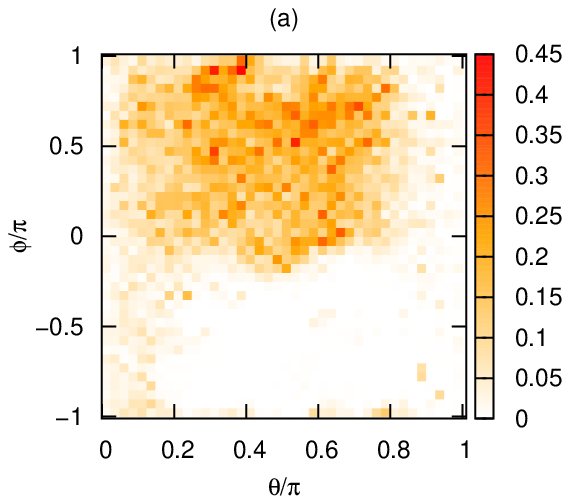}
\end{minipage}
\hspace{1cm}
\begin{minipage}{.42\linewidth}
\includegraphics[scale=1.2]{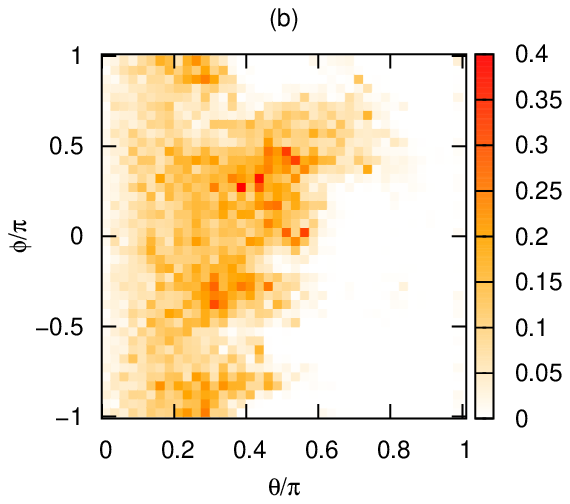}
\end{minipage}
\begin{minipage}{.42\linewidth}
\includegraphics[scale=1.2]{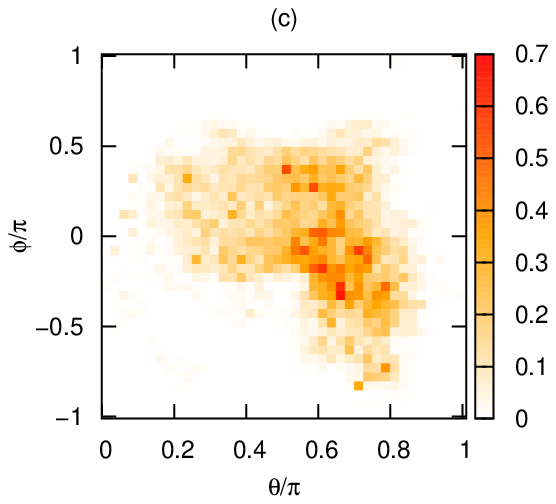}
\end{minipage}
\hspace{1cm}
\begin{minipage}{.42\linewidth}
\includegraphics[scale=1.2]{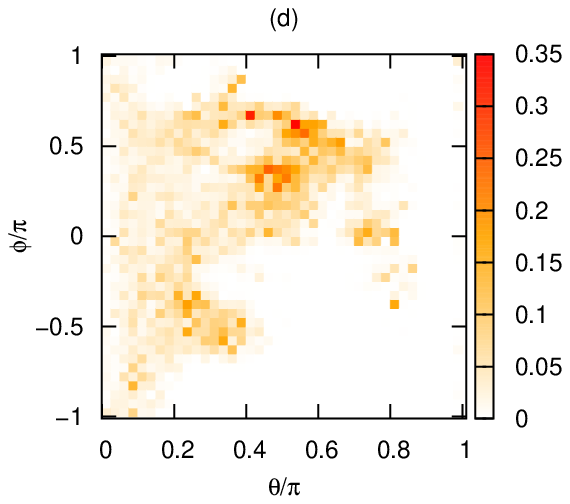}
\end{minipage}
\end{center}
\caption{
Angular excitation functions for the resonant normal modes
of water
for the (a)  81st, (b) 84th, (c) 93rd, and (d) 97th residues, 
represented in arbitrary units.
}
\label{fig:dist2-wat}
\end{figure}

Finally we examine the angular excitation functions
in Figs.~\ref{fig:dist2-pro} and \ref{fig:dist2-wat} 
to understand the VER pathways along the angular directions. 
We observe that the VER of the amide I modes is spatially anisotropic.
For the protein (Fig.~\ref{fig:dist2-pro}), 
the resonant modes 
are mainly localized around $\phi=0$ or $\pm \pi$,
which indicates that VER occurs in the peptide backbone plane 
(see Fig.~\ref{fig:geometry}).
For the 81st and 84th residues, belonging to a loop
structure, the value around $\phi \sim 0$ enhances 
because the amide plane is planer and the intramolecular 
contribution comes from such a direction.
For the 93rd and 97th residues, belonging to 
an $\alpha$ helical structure, 
the value around $\phi \sim 0$ diminishes
because the amide I plane is bended.

On the other hand, for the water (Fig.~\ref{fig:dist2-wat}), 
the angular excitation function is rather broad, and 
these distributions are similar to the conventional 
distribution function of water around the CO bond (not shown here). 
However, the excitation functions are less uniform compared 
to the conventional distribution function of water, implying 
the anisotropy of VER into water.
While these excitation functions can not be 
directly measured, it is of significant interest to examine 
the anisotropy of energy flow in proteins
using experimental methods such as the multidimensional 
spectroscopy \cite{ZH01,WH02,HLH98,ZAH01,MKFKAZ04}.

Recently Dlott and coworkers \cite{Dlott02} and 
Hamm and coworkers \cite{Hamm07} 
devised a new experimental technique to clarify energy transfer pathways 
in molecular systems. This method may be also combined with our analysis 
to clarify the water contribution of VER in a protein.

\section{Concluding remarks}
\label{sec:summary}

Using the time-dependent perturbation formula 
we developed in our previous paper,
we have investigated the vibrational energy relaxation (VER) 
of amide I modes in cytochrome c solvated with water.
We observed that the VER rates are sub-picosecond 
for the four residues we examined here, which are in
accord with previous experiment for other proteins.
We have decomposed VER into separate contributions from 
protein and water, and further projected the resonant modes 
onto radial and angular excitation functions.
Although the total VER rate is similar, 
the detailed mechanism of VER is different; 
there are protein-mediated pathways and water-mediated pathways
depending on which residue is excited.
There is anisotropy of VER for the water-mediated pathways
as well as for the protein-mediated pathways,
implying experimental consequences by multidimensional 
spectroscopy or other means.


Anharmonicity is a key issue in VER calculations.
We note that our simulated dynamics are based on 
the CHARMM force field, in which 
vibrational anharmonicity has been underestimated.
There is some evidence
that the anharmonic coupling calculated using the force field 
can be comparable to that derived from {\it ab initio} calculations
(Fujisaki, H.; Yagi, K.; Hirao, K.; Straub, J.E. unpublished).
However, without such a direct comparison,
it is unclear how accurate the results of our force field calculations are.
Nevertheless, the use of empirical energy functions is currently the only 
feasible way to characterize VER in large molecules.
To expand on these studies, we must develop a new methodology to investigate 
the quantum dynamics of the amide I mode with {\it ab initio}
potential surfaces. 
As first steps, we investigated N-methylacetamide in 
vacuum using the VCI method \cite{FYHS07} and in water cluster 
using the perturbation method (Zhang, Y.; Fujisaki, H.; Straub, J.E. unpublished) 
on {\it ab initio} potentials.
How to extend these studies to larger systems using 
multiresolution methods \cite{rauhut,YHH06} or QM/MM methods 
\cite{LSCLS06,Hirata05,ST07}
would be our focus in the near future.

Another important and related topic which should be pursued by such {\it ab initio} methods
is the effect of 
polarization \cite{HHLKKC04,Skinnergroup,ZAHM06,HJZM05,Stock06}. 
Morita and Kato clarified that polarization is 
important for VER of a stretching mode in azide ion in water \cite{MK98}, 
and this might be also the case for VER of amide I modes in a protein.
In particular, it is important to investigate 
some portion of a protein where the internal electric field 
is large and the polarization effect is expected to be enhanced. 

\acknowledgments

We thank Y.~Zhang for a useful suggestion on the relation between the VER rate and 
SASA and anonymous referees for useful suggestions. We are also grateful to 
Prof.~G.~Stock, Prof.~D.M.~Leitner, Prof.~K.~Hirao, 
Prof.~B.~Brutschy, Prof.~J.~Wachtveitl,
Dr.~K.~Yagi, Dr.~A.~Furuhama for useful discussions. 
We thank the National Science 
Foundation (CHE-0316551), Boston University's Center for Computer Science 
for generous support to our research, and 
the Department of Chemistry and Biochemistry at Montana State University
for generous support and hospitality.





\begin{thebibliography}{00}




%

\bibitem{KB86}
Krimm, S.; Bandekar, J.
{\it Adv.~Prot.~Chem.~}{\bf 1986}, {\it 38}, 181.

\bibitem{TT92}
(a) Torii, H.; Tasumi, M. 
{\it J. Chem. Phys.} {\bf 1992}, {\it 96}, 3379.
(b) Torii, H.; Tasumi, M. 
{\it J. Chem. Phys.} {\bf 1992}, {\it 97}, 92.

\bibitem{BZ02}
Barth, A.; Zscherp, C.
{\it Q.~Rev.~Biophys.~}{\bf 2002}, {\it 35}, 369.

\bibitem{SAD03}
Silva, R.A.G.D.;
Barber-Armstrong, W.; 
Decatur, S.M.
{\it J.~Am.~Chem.~Soc.~}{\bf 2003}, {\it 125}, 13674.

\bibitem{GHG04}
Gnanakaran, S.; Hochstrasser, R.M.; Garcia, A.E.
{\it Proc.~Natl.~Acad.~Sci.~U.S.A.} {\bf 2004}, {\it 101}, 9229.


\bibitem{TM93}
(a) Tanimura, Y.; Mukamel, S.
{\it J. Chem. Phys.} {\bf 1993}, {\it 99}, 9496.
(b) Mukamel, S.
{\it Principles of Nonlinear Optical Spectroscopy} 
(Oxford Univ. Press, New York, 1995).


\bibitem{ZH01}
Zanni, M.T.; Hochstrasser, R.M.
{\it Curr.~Opin.~Struc.~Biol.~}{\bf 2001}, {\it 11}, 516.

\bibitem{WH02}
Woutersen, S.; Hamm, P. 
{\it J.~Phys.~Cond.~Matt.~}{\bf 2002}, {\it 14}, 1035.

\bibitem{WH06}
Wang, J.P.; Hochstrasser, R.M.
{\it J.~Phys.~Chem.~B} {\bf 2006}, {\it 110}, 3798.

\bibitem{Tokmakoff06}
DeFlores, L.P.; Ganim, Z.; Ackley, S.F.; Chung, H.S.; Tokmakoff, A.
{\it J.~Phys.~Chem.~B} {\bf 2006}, {\it 110} 18973.


\bibitem{IT06}
Ishizaki, A.; Tanimura, Y.
{\it J. Chem. Phys.} {\bf 2006}, {\it 125}, 084501.

\bibitem{NTM07}
Nagata, Y.; Tanimura, Y.; Mukamel, S.
{\it J. Chem. Phys.} {\bf 2007}, {\it 126}, 204703.

\bibitem{Torii07}
Torii, H.
{\it J. Phys. Chem. B} {\bf 2007}, {\it 111}, 5434.


\bibitem{HHLKKC04}
(a) Ham, S.; Hahn, S.; Lee, C.; Kim, T.K.;  Kwak, K.; Cho, M.
{\it J. Phys. Chem. B} {\bf 2004}, {\it 108}, 9333;
%
(b) Ham, S.; Hahn, S.; Lee, C.; Cho, M.
{\it J. Phys. Chem. B} {\bf 2005}, {\it 109}, 11789.

\bibitem{Skinnergroup}
(a) Schmidt, J.R.; Corcelli, S.A.; Skinner, J.L. 
{\it J. Chem. Phys.} {\bf 2004}, {\it 121}, 8887. 
(b) Li, S.; Schmidt, J.R.; Corcelli, S.A.; Lawrence, C.P.; Skinner, J.L. 
{\it J. Chem. Phys.} {\bf 2006}, {\it 124}, 204110.

\bibitem{ZAHM06}
Zhuang, W.; Abramavicius, D.; Hayashi, T.; Mukamel, S. 
{\it J. Phys. Chem. B} {\bf 2006}, {\it 110}, 3362.
%

\bibitem{HJZM05}
Hayashi, T.; la Cour Jansen, T.; Zhuang, W.; Mukamel, S. 
{\it J. Phys. Chem. A} {\bf 2005}, {\it 109}, 64. 



\bibitem{Stock06}
Gorbunov, R.D.; Nguyen, P.H.; Kobus, M.; Stock, G.
{\it J.~Chem.~Phys.~} {\bf 2007}, {\it 126}, 054509.

\bibitem{GCG02}
Gregurick, S.K.; Chaban, G.M.; Gerber, R.B. 
{\it J.~Phys.~Chem.~A} {\bf 2002}, {\it 106}, 8696. 

\bibitem{BS06}
Bounouar, M.; Scheurer, Ch.
{\it Chem.~Phys.~} {\bf 2006}, {\it 323}, 87.

\bibitem{KB07}
Kaledin, A. L.; Bowman, J. M.
J. Phys. Chem. A. {\bf 2007}, {\it 111}, 5593-5598, 
doi:10.1021/jp0723822.

\bibitem{FYHS07}
Fujisaki, H.; Yagi, K.; Hirao, K.; Straub, J.E.
{\it Chem.~Phys.~Lett.} {\bf 2007}, {\it 443}, 6-11, 
doi:10.1016/j.cplett.2007.06.067, e-print arXiv:0706.1905. 


\bibitem{HLH98}
Hamm, P.; Lim, M.H.; Hochstrasser, R.M.;
{\it J.~Phys.~Chem.~B} {\bf 1998}, {\it 102}, 6123. 

\bibitem{ZAH01}
Zanni, M.T.; Asplund, M.C.; Hochstrasser, R.M.
{\it J.~Chem.~Phys.~} {\bf 2001}, {\it 114}, 4579. 


\bibitem{XMHA00}
(a) 
Xie, A.; van der Meer, L.; Hoff, W.; Austin, R.H.
{\it Phys.~Rev.~Lett.~} {\bf 2000}, {\it 84}, 5435.
(b) 
Austin, R.H.; Xie, A.; van der Meer, L.; Redlich, B.; Lindgard, P.A.; 
Frauenfelder, H.
{\it Phys.~Rev.~Lett.~} {\bf 2005}, {\it 94}, 128101.



\bibitem{NS03}
Nguyen, P.H.; Stock, G.
{\it J.~Chem.~Phys.~} {\bf 2003}, {\it 119}, 11350.



\bibitem{FZS06}
Fujisaki, H.; Zhang, Y.; Straub, J.E.
{\it J.~Chem.~Phys.~} {\bf 2006}, {\it 124}, 144910.


\bibitem{FBS05}
(a) 
Fujisaki, H.; Bu, L.; Straub, J.E.
{\it Adv. Chem. Phys. } {\bf 2005}, {\it 130B}, 179.
(b)
Fujisaki, H.; Bu, L.; Straub, J.E.
%
in {\it Normal Mode Analysis: Theory and Applications to Biological and Chemical Systems}, 
edited by Q. Cui and I. Bahar, Chapman and Hall/CRC Press, Boca Raton, Florida (2005).
(c)
Fujisaki, H.; Straub, J.E.
{\it Proc.~Natl.~Acad.~Sci.~U.S.A.} {\bf 2005}, {\it 102}, 6726.

\bibitem{Cremeens06}
Cremeens, M.; Fujisaki, H.; Zhang, Y.; Zimmermann, J.; Sagle, L.B.; 
Matsuda, S.; Dawson, P.E.; Straub, J.E.; Romesberg, F.E.;
{\it J. Am. Chem. Soc.} {\bf 2006}, {\it 128}, 6028. 




\bibitem{MKFKAZ04}
(a) 
Mukherjee, P.; Krummel, A.T.; Fulmer, E.C.; Kass, I.; Arkin, I.T.; Zanni, M.T.
{\it J.~Chem.~Phys.~} {\bf 2004}, {\it 120}, 10215.
(b)
Mukherjee, P.; Kass, I.; Arkin, I.T.; Zanni, M.T.
{\it J.~Phys.~Chem.~B} {\bf 2006}, {\it 110}, 24740.


\bibitem{Arkin06}
Arkin, I.T. 
{\it Curr.~Opin.~Chem.~Biol.~} {\bf 2006}, {\it 10}, 394.



\bibitem{YHTSG04}
Yagi, K.; Hirao, K.; Taketsugu, T.; Schmidt, M.W.; Gordon, M.S.
{\it J.~Chem.~Phys.~} {\bf 2004}, {\it 121}, 1383.


\bibitem{SO98}
(a) Shiga, M.; Okazaki, S.
{\it J.~Chem.~Phys.~} {\bf 1998}, {\it 109}, 3542.
(b) Shiga, M.; Okazaki, S.
{\it J.~Chem.~Phys.~} {\bf 1999}, {\it 111}, 5390.
(c)
Mikami, T.; Shiga, M.; Okazaki, S.
{\it J.~Chem.~Phys.~} {\bf 2001}, {\it 115}, 9797.
(d)
Mikami, T.; Okazaki, S.
{\it J.~Chem.~Phys.~} {\bf 2004}, {\it 121}, 10052.


\bibitem{Leitner05}
(a) Yu, X.; Leitner, D.M.
{\it J.~Phys.~Chem.~B} {\bf 2003}, {\it 107}, 1698.
(b) Leitner, D.M. 
{\it Adv.~Chem.~Phys.~} {\bf 2005}, {\it 130B}, 205.
(c) Leitner, D.M.; Havenith, M.; Gruebele, M.
{\it Int.~Rev.~Phys.~Chem.~} {\bf 2006}, {\it 25}, 553.


\bibitem{DK05}
Dijkstra, A.G.; Knoester, J.
{\it J.~Phys.~Chem.~B} {\bf 2005}, {\it 109}, 9787.


\bibitem{SS01}
Sagnella, D.E.; Straub, J.E.
{\it J.~Phys.~Chem.~B} {\bf 2001}, {\it 105}, 7057.

\bibitem{ON05}
Okamoto, T.; Nagaoka, M. 
{\it Chem.~Phys.~Lett.~} {\bf 2005}, {\it 407}, 444.

\bibitem{MO03}
Mikami, T.; Okazaki, S. 
{\it J.~Chem.~Phys.~} {\bf 2003}, {\it 119}, 4790. 



\bibitem{CHARMM}
(a) Brooks, B.R.; Bruccoleri, R.E.; Olafson, B.D.; 
States, D.J.; Swaminathan, S.; Karplus, M.
{\it J.~Comp.~Chem.~} {\bf 1983}, {\it 4}, 187.
(b)
MacKerell, Jr. A.D.; Brooks, B.; Brooks III, C.L.; Nilsson, L.; Roux, B.;
Won, Y.; Karplus, M.
in {\it The Encyclopedia of Computational Chemistry}, {\it 1}, 271, 
edited by P.v.R.~Schleyer {\it et al}., John Wiley \& Sons: Chichester (1998).

\bibitem{BS03}
Bu, L.; Straub, J.E.
{\it Biophys.~J.~} {\bf 2003}, {\it 85}, 1429.


\bibitem{Stratt95}
(a) Stratt, R.M. 
{\it Acc.~Chem.~Res.~} {\bf 1995}, {\it 28}, 201.
(b) Keyes, T. 
{\it J.~Phys.~Chem.~A} {\bf 1997}, {\it 101}, 2921. 

\bibitem{ZFS07}
Zhang, Y.; Fujisaki, H.; Straub, J.E.
{\it J.~Phys.~Chem.~B} {\bf 2007}, {\it 111}, 3243. 

\bibitem{MMK00}
(a) Moritsugu, K.; Miyashita, O.; Kidera, A.
{\it Phys.~Rev.~Lett.~} {\bf 2000}, {\it 85}, 3970.
(b) 
Moritsugu, K.; Miyashita, O.; Kidera, A.
{\it J.~Phys.~Chem.~B} {\bf 2003}, {\it 107}, 3309.


\bibitem{Dlott02}
Wang, Z.; Pakoulev, A.; Dlott, D.D.
{\it Science} {\bf 2002}, {\it 296}, 2201.

\bibitem{Hamm07}
Botan, V; Backus, E.H.G.; Pfister, R.; Moretto, A.;
Toniolo, C.; Nguyen, P.H.; Stock, G.; Hamm, P.
{\it Proc.~Natl.~Acad.~Sci.~U.S.A.} {\bf 2007}, (in press).




\bibitem{rauhut}
Rauhut, G. 
{\it J.~Chem.~Phys.} {\bf 2004}, {\it 121} 9313.

\bibitem{YHH06}
Yagi, K.; Hirata, S.; Hirao, K.
{\it Theo.~Chem.~Acc.} {\bf 2007}, (in press), doi 10.1007/s00214-007-0363-x.


\bibitem{LSCLS06}
Lin, S.Z.; Schmidt, J.R.; Corcelli, S.; Lawrence, C.P.; Skinner, J.L.;
{\it J.~Chem.~Phys.} {\bf 2006}, {\it 124}, 204110.

\bibitem{Hirata05}
Hirata, S.; Valiev, M.; Dupuis, M.; Xantheas, S.S.; Sugiki, S.; Sekino, H.
{\it Mol.~Phys.} {\bf 2005}, {\it 103} 2255.

\bibitem{ST07}
Shiga, M.; Tachikawa, M.
{\it Mol. Sim.} {\bf 2007}, {\it 33}, 171.

\bibitem{MK98} 
(a) Morita, A.; Kato, S. 
{\it J. Chem. Phys.} {\bf 1998}, {\it 109}, 5511.
(b)
Li, S.; Schmidt, J.R.; Skinner, J.L. 
{\it J. Chem. Phys.} {\bf 2006}, {\it 125}, 244507.

\bibitem{VMD}
Humphrey, W.; Dalke, A.; Schulten, K.
{\it J.~Mol.~Graph.~} {\bf 1996}, {\it 14}, 33.





\end{thebibliography}
\end{document}